\documentclass[12pt]{article}
\usepackage{epsfig}
\usepackage{graphicx}
\usepackage{a4}
\usepackage{axodraw}
\usepackage{latexsym}
\usepackage{cite}
\usepackage{xspace}
\usepackage{url}

\newcommand{\form}{{\sc Form}\xspace}
\newcommand{\tform}{{\sc TForm}\xspace}
\newcommand{\parform}{Par{\sc Form}\xspace}

\begin{document}

\begin{titlepage}
\noindent
\hfill Nikhef 2012-004

\hfill TTP12-008

\hfill SFB/CPP-12-15

\hfill March 2012

%
\begin{center}
\Large
{\bf FORM version 4.0} \\
\vspace{1.5cm}
\large
J. Kuipers$^{\, a}$, T. Ueda$^{\, b}$, J.A.M. Vermaseren$^{\, a}$ 
and J. Vollinga$^{\, a}$\\
\vspace{1.2cm}
\normalsize
{\it $^a$Nikhef Theory Group \\
\vspace{0.1cm}
Science Park 105, 1098 XG Amsterdam, The Netherlands} \\
\vspace{0.5cm}
{\it $^b$Institut f\"ur Theoretische Teilchenphysik,
Karlsruhe Institute of Technology (KIT) \\
\vspace{0.1cm}
D-76128 Karlsruhe, Germany} \\
\vspace{2.0cm}
\large
{\bf Abstract}
\vspace{-0.2cm}
\end{center}

We present version 4.0 of the symbolic manipulation system \form. The
most important new features are manipulation of rational polynomials
and the factorization of expressions. Many other new functions and
commands are also added; some of them are very general, while others
are designed for building specific high level packages, such as one
for Gr\"obner bases. New is also the checkpoint facility, that allows
for periodic backups during long calculations. Lastly, \form 4.0 has
become available as open source under the GNU General Public License 
version 3.
\vspace{1.0cm}
\end{titlepage}


\section{Introduction}

Over the years the symbolic manipulation system
\form~\cite{Vermaseren:2000nd} has undergone many changes. When its
first version was released in 1989, it was capable of dealing with the
problems that existed at that time: taking traces of strings of Dirac
gamma matrices, executing operations on symbols and dot products of
vectors, and manipulating functions with simple arguments. As a
successor of Schoonschip~\cite{HStrubbe} \form went in most of these
properties just a bit beyond its predecessor's capabilities. Its main
features were already its speed and its facilities for very large
expressions.

Over the years the computations done with \form became more and more 
complex and hence the program was extended regularly. One of the focuses of 
development has been the dealing with functions, their argument fields and 
pattern matching. More communication between the contents of the 
expressions and the input program has been provided with introducing the 
\$-variables in version 3. This version also contained a complete rewriting 
of the system that processes the input, which made it much easier to add 
new commands. In addition the language became a bit more coherent. \form 
version 3 was the result of 11 years of experience with different users 
doing calculations and this had produced a system with many original 
features. However, one of the capabilities that was missed by many users 
was the factorization of expressions. This has been added in version 4. 
Another feature that was requested by many users was the ability to work 
with rational polynomials as coefficients of terms. In the past this 
problem was addressed by letting \form interact with external 
programs~\cite{extform} and send such polynomial work to other programs 
like Fermat~\cite{FERMAT}. After the external program has finished its 
work, \form would receive the results and proceeds from there. In version 4 
this is no longer necessary for rational polynomial arithmetic. 
Nonetheless, the external communications facility still remains and can be 
used for many other things.

Another feature, that has been added over the years, was running \form on 
more than a single processor. This started with the \parform~\cite{parform} 
project in Karlsruhe. \parform is developed to run across different 
computers over a network using MPI, but it can also be used on computers 
with multiple cores. For this last type of computers also a multithreaded 
version of \form was created, named \tform~\cite{tform}. In \tform most 
features were easier to implement. The disadvantage of \parform is that it 
has more communication overhead. The disadvantage of \tform when compared 
to \parform is that \tform can only run on single multicore computers. When 
all workers have a heavy interaction with the (usually single) disk, there 
may be a severe slowdown due to a traffic jam at the disk. When \parform 
runs over a network on several computers, each with their own disk, this 
problem does not occur. For both \parform and \tform the computing model is 
such that most programs that are created for regular (sequential) \form, 
will not need any modification to benefit from the extra cores by running 
faster. At times some statements can be added to the program to improve the 
performance even further.

One of the problems of developing a language is backward compatibility. The 
transition from version 2 to 3 saw a number of language features changed. 
To aid the users in the transition a conversion program was provided, but 
still there were cases in which manual intervention was necessary. Also, 
some users had been using, against the instructions of the manual, a 
variety of bugs in the system that were repaired in version 3. As a result 
the conversion of old version 2 code was sometimes a bit painful. The 
introduction of version 4 has none of these problems. All version 3 
programs should work with version 4. The only things that have been changed 
in this respect are some default settings, but they can be restored to 
their old values in the setup file if wanted.

With the introduction of version 4 some extra facilities become available. 
\form is now open source under the GNU General Public License\cite{GPL} 
version 3 and there is a publicly accessible CVS repository from which 
anybody can download the sources\footnote{The sources are accessible via:
\url{http://www.nikhef.nl/~form/formcvs.php}}.
In addition there is quite an amount of documentation. Another facility is 
the forum\footnote{The forum is accessible via:
  \url{http://www.nikhef.nl/~form/forum}}
on which users and developers can discuss with each other. This has already 
been very helpful in locating bugs and installation problems.

The outline of this paper is as follows. In section 2 we discuss the 
generic structure of \form, \tform and \parform. In section 3 we discuss 
\form being open source and address the problem of adding new features. 
Section 4 is dedicated to the new features that were introduced in the 
later editions of version 3 and in version 4. In section 5 details of the 
algorithms used for polynomial algebra are discussed. Some attention is 
given to the parallel versions in section 6. Section 7 addresses generic 
facilities and the conclusions are given in section 8.


\section{Generic structure}

Before going into the new features of version 4 it is best to consider 
first the internal structure of \form. The system consists of a number of 
modules that have a limited amount of interaction with each other, namely
\begin{itemize}
\item the preprocessor,
\item the compiler,
\item the pattern matcher,
\item the terms generator,
\item the normalization routines, and
\item the sorting system.
\end{itemize}
In addition there are routines that deal with specific tasks such as taking 
traces of Dirac gamma matrices, initialization, storage, and handling table 
bases. There are also libraries for many types of standard operations like 
arithmetic, writing files, and compression. When available, \form uses 
libz~\cite{zlib} for the gzip compression routines and the GNU Multiple 
Precision Arithmetic Library~\cite{GMP} for arithmetic. \tform uses the 
POSIX routines for multithreaded processing and \parform uses the MPI 
libraries for network communication.

The preprocessor reads the input and edits this input according to the 
preprocessor syntax. Each time a statement is completed it is sent to the 
compiler. Whenever an end-of-module instruction is encountered the combined 
compiled code in the compiler buffers is executed. Once all active 
expressions have been processed with this code, the compiler buffer is 
emptied and the preprocessor continues reading the input.

The compiler is offered complete statements and translates those first into 
an intermediate code which consists of so-called \emph{tokens}. At this 
intermediate level some optimizations are performed. Next the tokens are 
translated into machine code for a virtual machine. The code is placed 
inside a compiler buffer. During this operation identical subexpressions 
are recognized to avoid storing them more than once.

The pattern matcher is an extensive piece of code that matches the left 
hand sides of substitution statements, called \emph{patterns}, with the 
contents of the terms in the expressions. Its most complicated part is 
dealing with wildcard variables and backtracking when there are several 
choices for matching and the choices made so far do not lead to a match.

The center of the program is the terms generator. This is the core of the 
virtual machine. It controls the branching of the terms when substituting 
part of a term by one or more terms from either the compiler buffer, from 
other buffers (like tables, \$-variables or other expressions), or from the 
automatic generation of terms as can happen during the expansion of special 
functions.

The normalization routines bring terms to normal form. This is not 
completely trivial, because it also involves action to evaluate special 
functions.

Finally, the sort routines bring collections of terms into a standard 
representation so that they form a unique expression. Its special features 
are that it is in principle disk based and can deal with very large 
expressions in a relatively short time. Because the representation of the 
terms is also very compact, \form can work with very large expressions. For 
\tform and \parform the sorting system is extended, because with multiple 
workers involved the sorted results have to be combined into a single 
expression in the end.

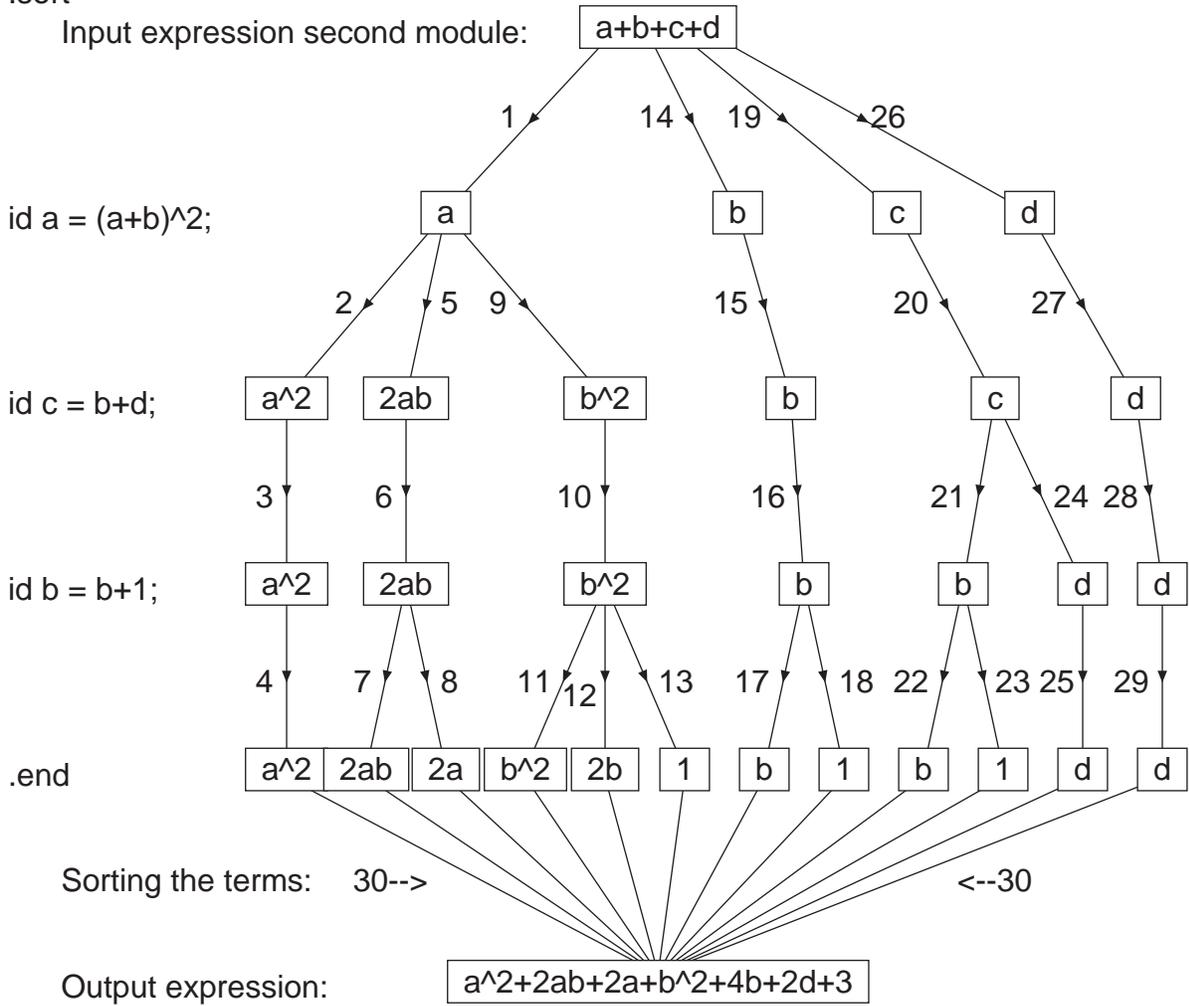
\begin{figure}
\centering
\begin{picture}(460,430)(0,0)
\SetPFont{Helvetica}{12}
\SetOffset(0,70)
\PText(5,350)(0)[l]{S   a,b,c,d;}
\PText(5,335)(0)[l]{On  HighFirst;}
\PText(5,320)(0)[l]{L	F = a+b+c+d;}
\PText(5,305)(0)[l]{.sort}
\PText(25,290)(0)[l]{Input expression second module:}
\ArrowLine(235,290)(170,220) \PText(197,258)(0)[r]{1}
\ArrowLine(245,290)(280,220) \PText(256,258)(0)[r]{14}
\ArrowLine(255,290)(340,220) \PText(289,258)(0)[r]{19}
\ArrowLine(265,290)(390,220) \PText(330,258)(0)[l]{26}
\BText(250,290){a+b+c+d}
\PText(5,220)(0)[l]{id a = (a+b)^2;}
\ArrowLine(170,220)(110,150) \PText(135,188)(0)[r]{2}
\ArrowLine(170,220)(155,150) \PText(168,188)(0)[l]{5}
\ArrowLine(170,220)(230,150) \PText(193,188)(0)[r]{9}
\BText(170,220){a}
\ArrowLine(280,220)(300,150) \PText(284,188)(0)[r]{15}
\BText(280,220){b}
\ArrowLine(340,220)(377,150) \PText(352,188)(0)[r]{20}
\BText(340,220){c}
\ArrowLine(390,220)(430,150) \PText(404,188)(0)[r]{27}
\BText(390,220){d}
\PText(5,150)(0)[l]{id c = b+d;}
\ArrowLine(110,150)(110,80) \PText(105,115)(0)[r]{3}
\BText(110,150){a^2}
\ArrowLine(155,150)(155,80) \PText(150,115)(0)[r]{6}
\BText(155,150){2ab}
\ArrowLine(230,150)(230,80) \PText(225,115)(0)[r]{10}
\BText(230,150){b^2}
\ArrowLine(300,150)(305,80) \PText(298,115)(0)[r]{16}
\BText(300,150){b}
\ArrowLine(377,150)(365,80) \PText(366,115)(0)[r]{21}
\ArrowLine(377,150)(410,80) \PText(399,115)(0)[l]{24}
\BText(377,150){c}
\ArrowLine(430,150)(440,80) \PText(431,115)(0)[r]{28}
\BText(430,150){d}
\PText(5,80)(0)[l]{id b = b+1;}
\ArrowLine(110,80)(110,10) \PText(105,45)(0)[r]{4}
\BText(110,80){a^2}
\ArrowLine(155,80)(140,10) \PText(142,45)(0)[r]{7}
\ArrowLine(155,80)(170,10) \PText(168,45)(0)[l]{8}
\BText(155,80){2ab}
\ArrowLine(230,80)(200,10) \PText(210,45)(0)[r]{11}
\ArrowLine(230,80)(230,10) \PText(227,40)(0)[r]{12}
\ArrowLine(230,80)(260,10) \PText(250,45)(0)[l]{13}
\BText(230,80){b^2}
\ArrowLine(305,80)(290,10) \PText(292,45)(0)[r]{17}
\ArrowLine(305,80)(320,10) \PText(318,45)(0)[l]{18}
\BText(305,80){b}
\ArrowLine(365,80)(350,10) \PText(352,45)(0)[r]{22}
\ArrowLine(365,80)(380,10) \PText(377,45)(0)[l]{23}
\BText(365,80){b}
\ArrowLine(410,80)(410,10) \PText(407,45)(0)[r]{25}
\BText(410,80){d}
\ArrowLine(440,80)(440,10) \PText(435,45)(0)[r]{29}
\BText(440,80){d}
\PText(5,10)(0)[l]{.end}
\Line(110,7)(250,-67)
\BText(110,10){a^2}
\Line(140,7)(250,-67)
\BText(140,10){2ab}
\Line(170,7)(250,-67)
\BText(170,10){2a}
\Line(200,7)(250,-67)
\BText(200,10){b^2}
\Line(230,7)(250,-67)
\BText(230,10){2b}
\Line(260,7)(250,-67)
\BText(260,10){1}
\Line(290,7)(250,-67)
\BText(290,10){b}
\Line(320,7)(250,-67)
\BText(320,10){1}
\Line(350,7)(250,-67)
\BText(350,10){b}
\Line(380,7)(250,-67)
\BText(380,10){1}
\Line(410,7)(250,-67)
\BText(410,10){d}
\Line(440,7)(250,-67)
\BText(440,10){d}
\PText(25,-30)(0)[l]{Sorting the terms:}
\PText(135,-30)(0)[l]{30-->}
\PText(391,-30)(0)[r]{<--30}
\PText(25,-70)(0)[l]{Output expression:}
\BText(250,-70){a^2+2ab+2a+b^2+4b+2d+3}
\end{picture} \vspace{2mm} \\
\caption{The tree expansion of terms in \form}
\label{fig:tree}
\end{figure}

\form's main objects are expressions. An expression is a sequence of terms 
stored sequentially on disk unless the expression is small enough to fit 
inside a cache buffer. The size of this buffer can be set by the user 
during startup. Each term is self-contained and contains no pointers. 
During execution no attempt is made to look for common subexpressions in 
different terms, hence there is no need for reference counts. In principle 
the terms are processed sequentially. This means that one term is taken 
from the input expression, the terms generator routine is called and it 
executes the first statement on it. The first term of the output of this 
statement is then processed by the generator routine with the second 
statement and so on. Figure \ref{fig:tree} shows this process.
%
%

After each statement the current term is normalized to make it ready for 
the next statement. When there are no more statements, the term is sent to 
the sorting system and the generator routine goes back one level, picks up 
the next term on that level, and then executes the next statement on it. 
This continues until the whole expansion tree has been traversed. The sort 
routines act semi-independently; they accept terms, put them in buffers, 
sort the buffers when full, and, if needed, write information to disk. 
After the last term has been sent to the sorting routines, the sorting is 
finalized. If the expression is big, this may involve disk-to-disk sorting. 
As soon as the input is no longer needed, it is destroyed and the output of 
the current module becomes the input for the next module.

Because all regular operations obtain their input from only a single term 
they are called local operations. Operations that take input from more than 
a single term are called non-local. Examples of non-local operations are 
the compares between terms during the sorting and the factorization of 
complete expressions. \form also has some semi-local operations. One of 
them is the \texttt{collect} statement. It places a predefined group of 
terms inside the argument of a function, which becomes part of a single 
term. Others involve the manipulation of functions of which one or more 
arguments contain expressions of more than a single term. When we will look 
at the polynomial operations we will see that they are inherently non-local 
although a large class of them can be treated as semi-local. The art of 
making good \form programs is often defined by how to construct inherently 
non-local operations by a combination of local operations and non-local 
sort instructions. We do not know of any systematic research of this.


\section{Open source and adding new features}

Starting with version 4, \form has become an open source program under the 
GNU General Public License~\cite{GPL} (GPL) version 3. The idea is to allow 
others to contribute to \form as well. Therefore it is important to 
describe shortly how new features can be added to it.

Adding new functions is easiest. There is a header file 
(\texttt{inivar.h}) in which all built-in functions are defined and 
one needs to add one line there with the name of the function and its 
internal number which is, given as a name, defined in another header 
file (\texttt{ftypes.h}). Then, depending on the function, its action 
has to be programmed. This is usually done in the normalization 
routine. In complicated cases some special routines may have to be 
added that are called either from the normalization routine or from 
the generator routine. The source code provides enough examples to 
illustrate this process.

Adding statements is often relatively simple in \form. The compiler 
has to recognize the keyword and send control to the routine that 
translates the statement. This routine has to be defined and it has to 
create the proper code for the virtual machine. The generator routine 
has to recognize this code and give control to the routine that 
executes the statement. Hence, apart from a few lines of code in other 
places only two routines have to be constructed.

Adding extra features in the \$-variables is similar, but requires 
more work.


\section{New features}

During the development of \form 3 many new features have been added. Most 
of those have been described before and are counted as belonging to version 
3. However, the ones that were added during the later stages of version 3.3 
are actually considered part of version 4 and we will describe them here. 
For a complete overview of all features one should consult the reference 
manual.


\subsection{Polynomial factorization}

In \form there are three different occurrences of polynomials, namely as 
arguments of functions, as complete expressions, and as \$-variables. These 
three cases are discussed separately.

\subsubsection{Function arguments}

When a polynomial is factorized, which is the argument of a function, the 
factors are returned as a sequence of arguments of the same function. The 
following program illustrates this.
\begin{verbatim}
    Symbols a,b,c;
    CFunction f;
    Format NoSpaces;
    Local F = f(3*(a+b)*(a+b+c)*(3*b+2*c));
    Print "<1> %t";
    FactArg f;
    Print "<2> %t";
    .end
<1> +f(6*b*c^2+15*b^2*c+9*b^3+6*a*c^2+21*a*b*c+18*a*b^2+6*a^2*c+9*a^2*b)
<2> +f(3,b+a,c+b+a,2*c+3*b)
\end{verbatim}
This use of the \texttt{FactArg} statement is different from what it used 
to be in version 3, when it would only find factors consisting of single 
terms. For backward compatibility we have provided the possibility to 
revert to the old behavior with the statement
\begin{verbatim}
    On OldFactArg;
\end{verbatim}
The new \texttt{FactArg} behavior is switched on again with
\begin{verbatim}
    Off OldFactArg;
\end{verbatim}
One may also specify the line
\begin{verbatim}
    OldFactArg ON
\end{verbatim}
in the setup file. This would be the easiest way to run old programs.

Dealing with polynomials as function arguments has the drawback that 
neither the input polynomial nor the combined output polynomials can be 
larger than the maximum term size.

\subsubsection{Expressions}

Expressions do not have the size constraint of function arguments, since 
they are disk based. However, their structure has no natural way to address 
the factors. To get around this we have selected a way that is in line with 
features of expressions: the bracket system.

First we adopt the convention that whether an expression is in factorized 
or unfactorized form is a choice that is made at the output level. The 
following program illustrates this.
\begin{verbatim}
    Symbols a,b,c;
    CFunction f;
    Format Nospaces;
    Local F = 6*b*c^3+15*b^2*c^2+9*b^3*c+6*a*c^3
             +21*a*b*c^2+18*a*b^2*c+6*a^2*c^2+9*a^2*b*c;
    Factorize F;
    Print;
    .end

Time =       0.00 sec    Generated terms =          8
               F         Terms in output =          8
                         Bytes used      =        312

Time =       0.00 sec    Generated terms =          9
               F         Terms in output =          9
               factorize Bytes used      =        368

   F=
       (3)
      *(c)
      *(c+b+a)
      *(2*c+3*b)
      *(b+a);
\end{verbatim}
The expression is first worked out as shown by the first statistics. Then 
the result is taken from the output and factorized. The factors are 
indicated by brackets in terms of the internal symbol \texttt{factor\_}. 
They are sorted as indicated by the second statistics. From this point on 
the expression is kept in factorized form, until it is unfactorized by 
using the statement \texttt{Unfactorize}. The internal representation of 
this factorized expression is
\begin{verbatim}
   F=
       factor_*(3)
      *factor_^2*(c)
      *factor_^3*(c+b+a)
      *factor_^4*(2*c+3*b)
      *factor_^5*(b+a);
\end{verbatim}
The writing routine does not write the special symbol \texttt{factor\_}, 
when the expression is in factorized form, but writes parentheses instead. 
It should be clear now that the substitution
\begin{verbatim}
   id  a^2 = 1;
\end{verbatim}
will have no effect on the factorized expression, since the left hand side 
of the \texttt{id} statement is not present in it.

When a substitution makes a factor zero, it is kept as a zero factor as 
shown by:
\begin{verbatim}
    Symbols a,b,c;
    CFunction f;
    Format Nospaces;
    Local F = 6*b*c^3+15*b^2*c^2+9*b^3*c+6*a*c^3
             +21*a*b*c^2+18*a*b^2*c+6*a^2*c^2+9*a^2*b*c;
    Factorize F;
    .sort
    id c = -b-a;
    Print;
    .end

   F=
       (3)
      *(-b-a)
      *(0)
      *(b-2*a)
      *(b+a);
\end{verbatim}
Unfactorizing the expression will result in zero, of course.
\begin{verbatim}
    Unfactorize F;
    Print;
    .end

   F=0;
\end{verbatim}
It is also possible to place factorized expressions directly into the 
input. For this the statements \texttt{LocalFactorized} and 
\texttt{GlobalFactorized} are present, which can be abbreviated as follows.
\begin{verbatim}
    Symbols a,b,c;
    CFunction f;
    LF F = 3*c*(a+b)*(a+b+c)*(3*b+2*c);
    Print;
    .end

   F =
         ( 3 )
       * ( c )
       * ( b + a )
       * ( c + b + a )
       * ( 2*c + 3*b );
\end{verbatim}
Factors are recognized by the multiplication or division signs at ground 
level, as shown here.
\begin{verbatim}
    Symbols a,b,c;
    CFunction f;
    LF F = 3/2*(3/2)*c*((a+b)*(a+b+c))*(3*b+2*c);
    Print;
    .end

   F =
         ( 3 )
       * ( 1/2 )
       * ( 3/2 )
       * ( c )
       * ( b*c + b^2 + a*c + 2*a*b + a^2 )
       * ( 2*c + 3*b );
\end{verbatim}
Once we know that the factors are stored as brackets it is easy to refer to 
them.
\begin{verbatim}
    Local G = F[factor_^3];
    Print G;
    .end

   G =
      3/2;
\end{verbatim}
The number of factors in factorized expression can be obtained by the 
function \texttt{NumFactors\_}.
\begin{verbatim}
    Local N = NumFactors_(F);
    Print N;
    .end

   N =
      6;
\end{verbatim}
A value of zero for the number of factors indicates that an expression is 
in the unfactorized state.

Extra \texttt{Bracket} statements will have no effect on factorized 
expressions. Currently \form supports only one level of brackets and this 
level is taken by the factorization. The brackets of the factorization use 
automatically the bracket index system to facilitate fast access to the 
brackets.

If one factorizes an already factorized expression, the system will try to 
factorize the factors. In addition it combines constant terms and the 
factors are sorted. If there are factors of zero the result will be zero, 
unless
\begin{verbatim}
    Factorize(keepzero) F;
\end{verbatim}
is used, in which case the zeroes are combined into a single zero as the 
first factor.

\subsubsection{\$-variables}

The third way of dealing with polynomials is as the content of 
\$-variables. This has yet other requirements, because the \$-variables 
have no access to the bracket system. In the case of the \$-variables we 
also have to take into account that they can be manipulated both at the 
preprocessor and the execution level. At the same time we have to realize 
that, although \$-variables are not restricted by the maximum term size, 
they do reside inside memory and therefore cannot be as large as regular 
expressions.

When a \$-variable has been factorized, we keep two copies of it: the 
unfactorized version and the factorized version. The factors remain in 
existence until the \$-variable is redefined.

The factorization is accomplished with the preprocessor instruction 
\texttt{\#FactDollar} or with the statement \texttt{FactDollar}. The 
factors are accessed by giving their number between square brackets after 
the name of the \$-variable as shown underneath. The zeroth element gives 
the number of factors. One may also use the \texttt{NumFactors\_} function 
to obtain the number of factors in a \$-variable. Again a zero return value 
indicates that the variable has not been factorized.
\begin{verbatim}
    Symbols a,b,c;
    #$X = (a+b+c)*3*(a+b)*(3*b+2*c);
    #FactDollar $X;
    #Message The first  factor of $X is `$X[1]'
~~~The first  factor of $X is 3
    #Message The second factor of $X is `$X[2]'
~~~The second factor of $X is c+b+a
    #Message The third  factor of $X is `$X[3]'
~~~The third  factor of $X is 2*c+3*b
    #Message The fourth factor of $X is `$X[4]'
~~~The fourth factor of $X is b+a
    #Message $X has `$X[0]' factors
~~~$X has 4 factors
    .end
\end{verbatim}
One may also use other \$-variables or factors of \$-variables between the 
brackets, provided they evaluate into valid factor numbers.
\begin{verbatim}
    Symbols a,b,c,x;
    CFunction f;
    Local F = f(3*(a+b)*(a+b+c)*(3*b+2*c))
             +f(2*(a+c)*(a+b+c));
    id  f(x?$x) = 1;
    FactDollar $x;
    do $i = 1,$x[0];
      Multiply f($x[$i]);
    enddo;
    Print +s;
    .end

   F =
       + f(2)*f(c + a)*f(c + b + a)
       + f(3)*f(b + a)*f(2*c + 3*b)*f(c + b + a)
      ;
\end{verbatim}
It should be noted that one cannot use constructions like
\begin{verbatim}
   Multiply <f($x[1])>*...*<f($x[$x[0]])>;
\end{verbatim}
because the triple dot operator is a preprocessor facility and
\verb:$x[0]: is not defined until the execution phase.


\subsection{Rational polynomials}

To speed up the manipulation of certain types of expressions \form is 
equipped with the \texttt{PolyFun} mechanism. This has made programs like 
Mincer~\cite{Mincer} considerably faster. There has been the need to extend 
this facility to dealing with rational polynomials. Therefore, the 
statement \texttt{PolyRatFun} is introduced in \form. It is declared in a 
similar way as the PolyFun:
\begin{verbatim}
   CFunction rat;
   PolyRatFun rat;
\end{verbatim}
The \texttt{PolyFun} and \texttt{PolyRatFun} declarations are mutually 
exclusive. The \texttt{PolyRatFun} is considered a special type of 
\texttt{PolyFun} and there can be only one of them at any moment. If one 
wants to switch back to a mode in which there is neither a \texttt{PolyFun} 
nor a \texttt{PolyRatFun} one can use
\begin{verbatim}
   PolyRatFun;
\end{verbatim}
to indicate that after this there is no function with that status.

In order for the \texttt{PolyRatFun} to be effective, the function needs 
two arguments and the arguments should contain symbols only. The first 
argument is then interpreted as the numerator of a fraction and the second 
as the denominator of that fraction. If there is only one argument it is 
seen as the numerator and \form will add a second argument that has the 
value one. The use of more than two arguments results in an error.

During the normalization of terms \form will work out products of more than 
one \texttt{PolyRatFun} as in:
\begin{verbatim}
    Symbols x,y;
    CFunction rat;
    PolyRatFun rat;
    Local F = rat(x,y)*rat(x+1,y+1);
    Print;
    .end

   F =
      rat(x^2 + x,y^2 + y);
\end{verbatim}
and during the sorting \form will add and subtract the fractions when the 
terms are otherwise identical as in
\begin{verbatim}
    Symbols x,y;
    CFunction rat;
    PolyRatFun rat;
    Local F = rat(x,y)+rat(x+1,y+1);
    Print;
    .end

   F =
      rat(2*x*y + x + y,y^2 + y);
\end{verbatim}
The fractions are at all times normalized. This means that the gcd of the 
numerator and the denominator has been divided out.

Often the denominator of a \texttt{PolyRatFun} can be factorized. This 
should however not be done while the function is still a PolyRatFun, 
because more than two arguments will cause an error. The following example 
shows a way of doing it:
\begin{verbatim}
    Symbols x,j,x1,x2;
    CFunction rat,num,den;
    PolyRatFun rat;
    Format NoSpaces;
    Local F = sum_(j,1,10,rat(1,x+j));
    Print;
    .sort

   F=
      rat(10*x^9+495*x^8+10560*x^7+127050*x^6+946638*x^5+4510275*x^4+
      13667720*x^3+25228500*x^2+25507152*x+10628640,x^10+55*x^9+1320*
      x^8+18150*x^7+157773*x^6+902055*x^5+3416930*x^4+8409500*x^3+
      12753576*x^2+10628640*x+3628800);

    PolyRatFun;
    id rat(x1?,x2?) = num(x1)*den(x2);
    FactArg num den;
    ChainOut,num;
    ChainOut,den;
    id num(x1?number_) = x1;
    id num(x1?symbol_) = x1;
    id den(x1?number_) = 1/x1;
    id den(x1?symbol_) = 1/x1;
    Print;
    .end

   F=
      num(11+2*x)*num(966240+2143152*x+1903836*x^2+896368*x^3+247049*
      x^4+41140*x^5+4070*x^6+220*x^7+5*x^8)*den(1+x)*den(2+x)*den(3+x)*
      den(4+x)*den(5+x)*den(6+x)*den(7+x)*den(8+x)*den(9+x)*den(10+x);
\end{verbatim}


\subsection{ToPolynomial, FromPolynomial and ExtraSymbols}

Sometimes it is better to have only symbols in an expression. This is for 
instance the case when an expression is put in the output for further 
processing in a C or Fortran program. The \texttt{ToPolynomial} statement 
converts all objects that are not symbols with positive powers into newly 
defined symbols and puts their definition in memory. The definitions of 
these symbols can also be printed. These symbols are called \emph{extra 
symbols} and will have names that do not interfere with any user defined 
names. Alternatively the user has control over their names and can 
influence whether they are printed as individual or as array elements. This 
last option can be rather handy in strong typed languages as C. An example 
of these statements is:
\begin{verbatim}
   Symbols x,y,z;
   CFunction f1,f2;
   ExtraSymbols array Ab;
   Local F = f1(x+y)*z+f2(z)*(x+y);
   ToPolynomial;
   .sort
   Format C;
   #write <example.c> "    int Ab[{`extrasymbols_'+1}],x,y,z;\n"
   #write <example.c> "%X"
   #write <example.c> "    F = %e",F
   .end
\end{verbatim}
This program produces the following output in the file \texttt{example.c}:
\begin{verbatim}
    int Ab[3],x,y,z;

    Ab[1]=f1(y + x);
    Ab[2]=f2(z);

    F = z*Ab[1] + y*Ab[2] + x*Ab[2];

\end{verbatim}
If the \texttt{ExtraSymbols} statement is omitted, the default names for 
the extra symbols are \texttt{Z1\_, Z2\_}, etc. If the \texttt{Array} 
option is omitted or the option \texttt{Underscore} is used the notation 
with the trailing underscore is used in the output, independent of the name 
selected. The statement \texttt{FromPolynomial} undoes the action of the 
\texttt{ToPolynomial} statement.

New features like factorization are implemented internally in terms of 
symbols only. Therefore the routines that execute the \texttt{ToPolynomial} 
and \texttt{FromPolynomial} statements are used also internally for the 
factorization and in the future they may be used for more features.


\subsection{Div\_, Rem\_, Gcd\_, Content\_ and Inverse\_}

Now that \form has polynomial algebra implemented internally, there is need 
for several functions, that are available to users, that can do some basic 
operations. These are:
\begin{description}
\item[Div\_] The quotient of the division of the first by the second 
argument. The arguments can be polynomials.
\item[Rem\_] The remainder of the division of the first by the second 
argument. The arguments can be polynomials.
\item[Gcd\_] The gcd of all the arguments. The arguments can be 
polynomials.
\item[Content\_] The content of the argument, i.e., the argument divided by 
the gcd of all terms.
\item[Inverse\_] The inverse of the first argument $P_1$ modulo the second 
argument $P_2$, i.e., more explicitly, the polygon $P_3$ such that 
$(P_3\cdot P_1) = 1$ mod $P_2$.
\end{description}
Functions together with their arguments are usually restricted by the 
condition that they have to fit inside the maximum size of a term. With 
regular functions and an argument that is an expression, the expression is 
first expanded and hence subject to this restriction. The first four 
functions in this category however do not suffer from this limitation, 
because the expressions are only expanded during the operation itself. The 
same holds for when the arguments are \$-variables.


\subsection{Do loops}

The \texttt{\#Do} instruction already exists since the original version of 
\form. This is a preprocessor feature that acts purely on the text of the 
input. Until recently its role at the execution level was taken over by the 
\texttt{Repeat}/\texttt{EndRepeat} construction, possibly in combination 
with \texttt{If} statements. With the advent of factorization and the need 
for organized access to the factors this became too complicated and we have 
implemented a proper \texttt{Do}/\texttt{EndDo} construction. The first 
question is of course what should serve as loop variable. For the original 
\texttt{\#Do} instruction we use preprocessor variables, but we do not have 
these available at execution time. However, we do have the \$-variables, 
which allows us to make the construction:
\begin{verbatim}
   Do $i = 1,5;
      id only x^$i = f(F[factor_^$i]);
   EndDo;
\end{verbatim}
The boundary parameters in the \texttt{Do} statement should be either 
integers, that can be stored inside one \form word, or \$-variables. They 
can also be \$-variables with factor indicators as in
\begin{verbatim}
   Do $i = 1,$e[0];
      Multiply f($e[$i]);
   EndDo;
\end{verbatim}


\subsection{The transform statement}

There are several categories of manipulations of functions with either 
complicated arguments or many simple arguments that take enormous amounts 
of time when handled by means of the pattern matcher. An example is the 
following, where arguments of zero are replaced by ones and vice versa.
\begin{verbatim}
    Symbols x;
    CFunction f,g;
    Local F = f(1,0,1,0,0,1,0,1);
    Multiply g;
    repeat id g(?a)*f(x?,?b) = g(?a,1-x)*f(?b);
    id	f*g(?a) = f(?a);
    Print;
    .end

   F =
      f(0,1,0,1,1,0,1,0);
\end{verbatim}
In this example the program has to use the repeat statement to go 8 times 
through the pattern matcher, plus a ninth time in the last \texttt{id} 
statement. The \texttt{Transform} statement is a whole group of operations 
in which the much simpler matching is done once and the rest is done 
internally by \form:
\begin{verbatim}
    Symbols x;
    CFunction f,g;
    Local F = f(1,0,1,0,0,1,0,1);
    Transform f replace(1,last)=(1,0,0,1);
    Print;
    .end

   F =
      f(0,1,0,1,1,0,1,0);
\end{verbatim}
Basically, the \texttt{Transform} statement provides a syntax to deal with 
large groups of arguments. Each transformation consists of a subkey, 
indicating its type, followed by arguments that are enclosed by 
parentheses. Currently, there are 10 different subkeys. After that specific 
information may follow. The proper syntax of the subkeys is in the manual, 
but below follows a short description of what is available. The first 
transformations are quite generally applicable, while the latter are 
custom-made for a project on Multiple Zeta Values\cite{MZV,MZV2}.
\begin{description}
\item[replace] Like the \texttt{replace\_} function, but allows also
  replacement of numbers.
\item[encode] Combines a range of numerical arguments into a single
  number over a specified base number.
\item[decode] The inverse of the encode subkey: a single argument is
  converted into a range of arguments.
\item[reverse] Reverses a range of arguments.
\item[cycle] Applies a cyclic permutation to a range of arguments.
\item[permute] Permutes arguments according to a given permutation.
\item[tosumnotation] There are two ways to characterize harmonic sums
  and harmonic polylogarithms. In the sum notation the indices are
  non-zero integers and in the integral notation there are only the
  numbers -1,0,1. In the conversion from integral to sum notation a
  zero adds one to the absolute value of the nonzero number to the
  right of it as in $0,0,0, -1\rightarrow -4$.
\item[tointegralnotation] The inverse of tosumnotation, see example
  below.
\item[islyndon] Tests whether the indicated range of arguments forms a
  Lyndon word\footnote{One definition of a Lyndon word is the unique
    minimal cyclic permutation of a number of objects.} according to
  the ordering of the arguments in \form. A yes and no argument tell
  what the main term should be multiplied by, when the answer is yes
  or no respectively.
\item[tolyndon] Will permute the given range in a cyclic manner until
  it is (if possible) a Lyndon word according to the ordering of the
  arguments in \form. Also here yes and no arguments should be
  specified.
\end{description}
It is possible to use more than one subkey in a single transform statement. 
This improves efficiency. This is illustrated in an example from a program 
that manipulates Multiple Zeta Values. To start off with, the original 
version was:
\begin{verbatim}
    Symbol x,x1,x2;
    CFunction H,H1;
    Local F = H(3,4,2,6,1,1,1,2);
    Print "<1> %t";
    Repeat id H(?a,x?!{0,1},?b) = H(?a,0,x-1,?b);
    Print "<2> %t";
    Multiply H1;
    Repeat id H(x?,?a)*H1(?b) = H(?a)*H1(?b,1-x);
    id  H*H1(?a) = H(?a);
    Print "<3> %t";
    Repeat id H(x1?,x2?,?a) = H(2*x1+x2,?a);
    Print "<4> %t";
    .end
<1>  + H(3,4,2,6,1,1,1,2)
<2>  + H(0,0,1,0,0,0,1,0,1,0,0,0,0,0,1,1,1,1,0,1)
<3>  + H(1,1,0,1,1,1,0,1,0,1,1,1,1,1,0,0,0,0,1,0)
<4>  + H(907202)
\end{verbatim}
In term of a single \texttt{transform} statement this becomes:
\begin{verbatim}
    CF  H;
    L   F = H(3,4,2,6,1,1,1,2);
    Transform H tointegralnotation(1,last),
                replace(1,last)=(0,1,1,0),
                encode(1,last):base=2;
    Print;
    .end

   F =
      H(907202);
\end{verbatim}
This last version is faster, more readable, easier to program and less 
prone to errors.


\subsection{Random\_ and RanPerm\_}

\texttt{Random\_} is a random number generator. It works according to an 
algorithm called the \emph{additive number generator} as described 
in~\cite{knuth}. The used subscript pair is $(38,89)$, which gives a longer 
cycle than the pair used in the reference. The function is called with an 
integer argument $N$.  This integer has to be greater than one. The return 
value is a random number in the range $[1,N]$. The generator can be 
initialized with the preprocessor instruction \texttt{\#SetRandom 
<number>}. If this instruction is never used the first use of the 
\texttt{Random\_} function will use a standard initialization. It should be 
noticed that in \tform and \parform each worker runs its own version of 
\texttt{Random\_} and they are initialized differently, even though this is 
controlled by the \texttt{\#SetRandom} instruction. One might argue that 
this makes the program give different results, depending on whether it is 
run with \form or with \tform, but that would be the case anyway, because 
the way the terms are distributed over the workers is non-deterministic. 
Hence, even if there were a unique sequence, the individual terms might not 
get the same number in the sequence.

An additional useful function is the \texttt{RanPerm\_} function. Its first 
argument is the name of the output function. It generates a random 
permutation of the remaining arguments and puts this permutation in the 
output function.
\begin{verbatim}
    CFunction f;
    Local F = RanPerm_(f,1,...,6);
    Print;
    .end
   F =
      f(3,1,6,2,4,5);
\end{verbatim}

These random functions can be helpful in testing and debugging programs 
among other things.


\subsection{FirstTerm\_}

The \texttt{FirstTerm\_} function needs a single argument that is the name 
of either an expression or a \$-variable. It returns the first term in that 
object. This can be useful in a Gaussian elimination scheme or in a program 
that computes a Gr\"obner basis. It is related to the 
\texttt{FirstBracket\_} function, which returns what is outside the first 
bracket of an expression that is in bracketed form.


\subsection{Prime\_, ExtEuclidean\_ and MakeRational\_}

Several types of calculations can be performed far more efficiently when 
working modulo prime numbers. Typically, results modulo different prime 
numbers can be combined with the Chinese remainder theorem to construct 
solutions over the integers or rationals. For this purpose, the functions 
\texttt{Prime\_}, \texttt{ExtEuclidean\_} and \texttt{MakeRational\_} are 
created.

The function \texttt{Prime\_} generates prime numbers, starting from the 
maximum positive number that fits inside a single \form word and working 
its way down. Numbers that are found are stored in a list, so that they do 
not have to be recomputed. The function \texttt{Prime\_(n)} gives the 
$n$-th element of that list. If it did not exist yet, \form has to compute 
it. Notice that this way of dealing with prime numbers allows only for 
constrained values of $n$. The minimum value of a prime number that is 
allowed is determined by the maximum value that a power of a symbol can 
have in \form. On a 64-bits computer this allows for ${\cal O}(10^8)$ prime 
numbers. One warning: the algorithm for prime number is not particularly 
fast and hence it might take of the order of a few hours to generate and 
store them all.

In the extended Euclidean algorithm one does not only determine the 
greatest common divisor $g$ of two numbers $n_1$ and $n_2$, but also two 
numbers $x_1$ and $x_2$ such that $g = x_1 n_1 + x_2 n_2$. In the case that 
$n_1$ and $n_2$ are relative prime (i.e., $g=1$) $x_1$ and $x_2$ are called 
the modular inverses of $n_1$ and $n_2$ because $x_1 n_1=1$ mod $n_2$ and 
so is $x_2 n_2$ mod $n_1$. This algorithm is implemented in \form in the 
function \texttt{ExtEuclidean\_}. It is useful for combining results modulo 
$n_1$ and modulo $n_2$ into a result modulo $n_1 n_2$ as the following 
example shows.
\begin{verbatim}
    #$p1 = Prime_(1);
    #$p2 = Prime_(2);
    Symbols x1,x2,x3,x4;
    Off Statistics;
    Local F = 12345678901234567;
    .sort
    Local G1 = Mod2_(F,$p1);
    Local G2 = Mod2_(F,$p2);
    .sort
    #$c1 = ExtEuclidean_($p1,$p2);
    #Inside $c1
        id ExtEuclidean_(x1?,x2?,x3?,x4?) = x2*x4;
    #EndInside
    #$c2 = ExtEuclidean_($p1,$p2);
    #Inside $c2
        id ExtEuclidean_(x1?,x2?,x3?,x4?) = x1*x3;
    #EndInside
    #$p3 = $p1*$p2;
    Modulus PlusMin `$p3';
    Local H = G1*$c1+G2*$c2;
    Print;
    .end

   F =
      12345678901234567;
   G1 =
       - 229487668;
   G2 =
       - 183496428;
   H =
      12345678901234567;
\end{verbatim}

\texttt{ExtEuclidean\_} always returns integer numbers, while often 
calculations result in rational numbers. To obtain fractions after 
calculating modulo integers, the function \texttt{MakeRational\_} is 
available. It takes two arguments, which both are integer. The function is 
replaced by the unique fraction of which both elements are less than the 
square root of the second argument and that, in calculus modulo this second 
number would give the same result as the first number modulo the second 
number. Example:
\begin{verbatim}
    #$m = prime_(1);
    #write <> "The prime number is %$",$m
The prime number is 2147483587
    Local F = MakeRational_(12345678,$m);
    Print;
    .sort

   F =
      9719/38790;

    Modulus `$m';
    Print;
    .end

   F =
      12345678;
\end{verbatim}


\subsection{Checkpoints}

There are various circumstances in which one can loose a large investment 
in computer time. One would be a power failure after many hours of running 
and another would be a syntax error near the end of a program. Whereas the 
second case is one the user can do something about by first testing the 
program on small examples, the first case is usually not under control. To 
avoid the potential loss of big computer resource investments, \form is 
equipped with a user controlled backup mechanism. At the end of a module the 
user has the possibility to make a snapshot of the current state. If, for 
one reason or another, the program fails at a later stage, it can be 
restarted from the last snapshot. This snapshot is stored in a disk file 
and contains the contents of all relevant files and all relevant contents 
of memory locations. It is tied to a given executable file of \form, \tform 
or \parform and only meant to continue execution when the failure condition 
has been resolved. One should not use it as a practical means of 
calculation backup. One should also not change the number of workers in 
\tform or \parform. \form remembers the reading position in open input 
files.  Hence, altering such a file before this position and changing its 
length may have disastrous consequences. Changing it after the current 
reading position should give no problems. This would be a way to repair an 
error that leads to a crash in the final stages of a program.

The checkpoint feature is activated and deactivated with the statements:
\begin{verbatim}
   On Checkpoint [<OPTIONS>];
   Off Checkpoint;
\end{verbatim}
The statement \texttt{On Checkpoint} will cause a snapshot to be made at 
the end of the module and each module that follows until a statement 
\texttt{Off Checkpoint} is encountered. If the checkpoints are already 
active and a new \texttt{On Checkpoint} statement is issued, the options of 
the new statement override the corresponding options of the old statement.

An option of the form
\begin{verbatim}
   <NUMBER>[<UNITS>]
\end{verbatim}
specifies a time interval and tells \form to do the writing at module ends 
only if more time than this given time has elapsed since the last recovery 
data write. The units are optional and default to seconds. Possible units 
are: {\tt s} for seconds, {\tt m} for minutes, {\tt h} for hours, and {\tt  
d} for days.

\form can run external programs before and after the recovery data writing. 
The options to specify the filename of such a program are:
\begin{verbatim}
   runbefore="<FILENAME>"
   runafter="<FILENAME>"
   run="<FILENAME>"
\end{verbatim}
The option {\tt run} specifies the filename for both cases simultaneously.  
{\tt <FILENAME>} needs to be a valid filename referring to a proper 
executable. Depending on the operating system this can mean the file needs 
to have the correct access permissions and to lie in the current search 
path. The programs will only be run if the condition on the time interval 
has been fulfilled. This can be used to check for enough disk space, and to 
compress or move the backups if desired.

The return value of the program run before the data writing will be 
interpreted. If it is a value unequal to zero or if the execution itself 
returns an error (e.g., if the executable cannot be found), no data will be 
written at this module's end. Independently, the program specified to be 
run after the data writing will always be executed.

The \texttt{-R} option in the calling of \form invokes the recovery 
procedure after a crash. Imagine that the original program was run with the 
command
\begin{verbatim}
   form -S my.set long.frm
\end{verbatim}
and it crashed, then
\begin{verbatim}
   form -S my.set -R long.frm
\end{verbatim}
will continue its execution at the point of the last snapshot.

The writing of the recovery data proceeds in several steps. First, all the 
data inside \form is written to an intermediate file. Second, existing 
files like a store or hide file are copied. At this point, the old recovery 
data basically gets overwritten. Finally, the intermediate data file is 
renamed into the proper recovery file. If any problem has occurred, it is 
signaled to the user. If during the copying of the store, hide, or scratch 
file anything goes wrong, the old recovery state will be overwritten and 
left in an unusable, undefined state. An improvement of the situation would 
be to copy all files first to intermediate files and finally rename them, 
but it would triple the amount of space needed on the disk. Because these 
files can be very large, this was considered impractical. The option 
\texttt{runbefore} can also be used to copy the backup first.


\subsection{System independent save files}

New is also the portability of the save files (files created with the 
\texttt{Save} statement and recovered with the \texttt{Load} statement) 
between computers with a different architecture. Of course the system does 
not work in 100\% of the cases, because on a computer with a 64-bits 
architecture one may define many more variables than on a computer with a 
32-bits architecture. If there are too many variables, the computer with 
the 32-bits architecture will give a nametable overflow when reading 
expressions with too many variables. Similarly there are different 
restrictions on maximum powers of symbols. The implementation of this 
portability made it necessary to redefine the format of the save files. 
Hence old save files will not work with version 4. The best way to convert 
old save files is by printing their contents to regular text output with 
the old version of \form and then reading the resulting expression with a 
new version of \form that can put it in a new save file.


\section{Polynomial algorithms}

For the manipulation of polynomials and the calculation of gcds and 
factorizations, \form uses a number of well-known algorithms. To store the 
polynomials, a degree sparse and variable dense representation is 
used~\cite{polyrep}. Degree sparse indicates that only non-zero terms are 
stored, so that the representation of sparse polynomials is efficient. 
Variable dense indicates that for each monomial a vector with the exponents 
of all variables is stored. This is slightly inefficient regarding storage, 
but typically speeds up hard polynomial computations such as factorization. 
For some algorithms, dense univariate polynomials are stored as an array of 
coefficients if that increases the efficiency.

Polynomial addition and subtraction is implemented by merging two 
polynomials. Multiplication and division is performed by using a binary 
heap of monomials to find the next term quickly and achieve good 
performance~\cite{polydiv}. All polynomial operations can be performed over 
the integers, modulo a prime number or modulo powers of primes. The latter 
are used for intermediate results in gcd computations and factorization. 
The prime powers are cached in a table for efficiency.

To calculate the gcd of two univariate polynomials, Euclid's algorithm is 
used~\cite{polyalgos}. For the multivariate case \form tries a heuristic 
algorithm a couple of times~\cite{polyheur}, which usually results in the 
answer. Upon failure, Zippel's sparse modular algorithm is 
used~\cite{polygcd}. This algorithm calculates univariate gcds modulo prime 
numbers and constructs from them the multivariate gcd with polynomial 
interpolation and the Chinese remainder algorithm.

For the factorization of univariate polynomials modulo prime numbers 
Berlekamp's algorithm is used~\cite{polyfact}. To obtain the factorization 
over the integers Hensel lifting is employed~\cite{polylift}. Eventual 
spurious factors are combined by a brute force approach. Multivariate 
polynomials are reduced to univariate ones by substituting appropriate 
integers for all but one variable, so that Berlekamp's algorithm can be 
used. The multivariate factorization is then constructed with the 
multivariate generalization of Hensel lifting. To circumvent as much of the 
multivariate lifting as possible, coefficients are predetermined by 
equating the factorization and the original polynomial~\cite{polypredet}. 
The leading coefficient problem is solved by Wang's 
algorithm~\cite{polypredet}.
 
A caching mechanism speeds up many realistic calculations in which the 
program attempts to factorize the same polynomials more than once.


\section{Parallel versions}

The fact that an expression in \form is a sequence of self-contained terms 
suggests that they can be processed in parallel for local operations. \form 
has two versions that allow the use of more than a single processor: 
\tform~\cite{tform} and \parform~\cite{parform}. \tform uses multiple 
threads in a single multicore computer in which the cores share the memory. 
\parform uses the MPI library to establish communication between different 
processes, which do not share the memory and may be on different computers, 
and to distribute the calculation over these processes. In principle both 
programs are supposed to work in the same way from the viewpoint of the 
user. In practice it was much easier to implement \tform, because one does 
not need to send information, each time some new data is needed. The cases 
that one needs to send data via the MPI on \parform include: 
\texttt{redefine} statements, \$-variables, expressions appearing in right 
hand sides of definition or substitution statements, and the global table 
for converting the extra symbols.

Both the parallel versions have been described before, but \parform has 
always been rather incomplete in the sense that some facilities were either 
missing or not working properly. With version 4 this has been fixed. It 
should now be possible to run all \tform programs also with \parform. Of 
course also all regular sequential \form programs should run both with 
\tform and \parform.

The two programs have different benefits. \tform needs far less 
communication and makes good use of several cores, but it suffers from the 
presence of only a single disk. \parform needs much more communication, but 
different computers usually have their own disk and hence there is less 
chance of slowdown due to traffic jams at the disk(s). It is up to the user 
to decide what is best for a given problem.

Both versions will suffer from bottlenecks. Some parts of the structure of 
\form cannot be parallelized in principle or are hard to parallelize. An 
example is the final sorting on the master. Hence there will be not much 
further improvement in the performance beyond the use of a given, problem 
dependent, number of processors. This is a problem that is worked on.


\section{Facilities}

With the upgrade to the new version also the manual has been extended. The 
complete manual comes in 4 versions:
\begin{description}
\item[.tex] the original LaTeX sources,
\item[.ps] the processed postscript file,
\item[.pdf] the processed PDF file, and
\item[.html] the online version.
\end{description}
This manual is a reference manual and should not be confused with a 
tutorial. The old tutorial by A.\ Heck and collaborators is still a very 
good introduction to \form, but it suffers from the fact that it was 
written for version 2. After version 2 the syntax has changed somewhat and 
also many new features were added. One of the future projects is to convert 
this tutorial to the syntax of version 4.

To make the open source aspect of \form more realistic new documentation 
has been written. It explains about the inner working of \form and many of 
its routines. Most routines have a header that is compatible with the 
documentation system Doxygen. This should help with finding ones way 
through the sources.

In addition a system has been programmed in the language Ruby to allow 
certification of new executables. In this system many examples of \form 
programs are stored, together with the crucial parts of their output.  
Making a certification run will test all these examples and whether they 
still give the correct answer. It is the intention to add more examples in 
the future.

To facilitate communication between users or between users and authors a 
forum has been created. Here people can report on new \form programs, 
problems with installation or execution, request new features, etc. We had 
to protect the forum a little bit, because at a given moment there were 
large numbers of spam attacks. Users should sign up and answer an easy 
question, before they are admitted and can make posts. Currently the forum 
can be found at \url{http://www.nikhef.nl/~form/forum}.

The \form program is all original code in the C and C++ languages. It uses 
a few generally available libraries. These are libgmp, librt, zlib, the 
POSIX multithread system, the MPI library, and, of course, the standard C 
and C++ libraries.

The GMP (GNU Multi Precision Arithmetic Library) deals with the arithmetic 
of very large numbers. Actually \form uses only the multiplication, the 
division and the gcd routines (all for integers) from this library. If the 
GMP library is not present, \form has its own routines which may be slower, 
because in the GMP library there are some assembler statements that do not 
have an equally efficient equivalent in the C language. In addition the GMP 
library uses better algorithms for extremely large numbers. The zlib is 
used for data (de)compression in the sort file (only if the user requests 
this) and for the tablebase facility. \form can be built without it. The 
POSIX library is used for \tform. Without it one cannot build \tform. The 
MPI library is used for \parform and without it one cannot build \parform.


\section{Conclusions}

The new version 4 brings a giant leap forward in the capabilities of \form. 
The facility that people missed most, factorization, has been added. Many 
new commands and functions pave the way for future calculations that apply 
significantly different algorithms from those used in the past. And if 
these features are not sufficient, users may implement their own additions.

{\bf Acknowledgments}: We are very grateful to the authors of zlib, the GMP 
library, the MPI library and the POSIX library to make their programs 
available in such an easy to use way.

The work of JK, JV and JV is part of the research programs of the 
"Stichting voor Fundamenteel Onderzoek der Materie (FOM)", which is 
financially supported by the "Nederlandse organisatie voor 
Wetenschappelijke Onderzoek (NWO)" and the work of TU is part of the 
research program of the DFG through SFB/TR 9 ``Computational Particle 
Physics''.

We would also like to thank the people who helped us with the debugging of 
the system. They spent much time in the preparation of good and 
comprehensible bug reports.


\end{document}